\documentstyle[12pt,epsfig]{article}
\newcommand{\be}{\begin{equation}}
\newcommand{\ee}{\end{equation}}
\newcommand{\ba}{\begin{eqnarray}}
\newcommand{\ea}{\end{eqnarray}}

\newcommand{\ga}{\gamma_5}
\newcommand{\dg}{^{\dagger}}
\newcommand{\re}[1]{(\ref{#1})}
\newcommand{\na}[1]{\nabla_{#1}}
\newcommand{\Id}{\mbox{1\hspace{-0.98mm}l}}   
\newcommand{\mb}[1]{\quad\mbox{ #1 }\quad}

\newcommand{\di}{\mbox{d}\,}

\begin{document}
\renewcommand{\baselinestretch}{1.1} \small\normalsize

\begin{flushleft}
\hfill {\sc HUB-EP}-99/50
\end{flushleft}

\vspace*{.9cm}

\begin{center}

{\Large \bf Differential equation for spectral flows\\
            of hermitean Wilson-Dirac operator}

\vspace*{0.8cm}

{\bf Werner Kerler}

\vspace*{0.3cm}

{\sl Institut f\"ur Physik, Humboldt-Universit\"at, D-10115 Berlin, 
Germany}
\hspace{3.6mm}

\end{center}

\vspace*{1.7cm}

\begin{abstract}
Establishing an exact relation for the derivative we show that the eigenvalue 
flows of the hermitean Wilson-Dirac operator obey a differential equation. We 
obtain a complete overview of the characteristic features of its solutions. 
The underlying mathematical aspects are fully clarified. 
\end{abstract}

\vspace*{.2cm}


\section{Introduction}

\hspace{3mm}
The hermitean Wilson-Dirac operator $H$ is a fundamental quantity in 
the overlap formalism \cite{na93,na95}. It has been shown there that the 
difference of the numbers of its positive and negative eigenvalues is related 
to the index of the massless Dirac operator. This gets particularly 
explicit with the Neuberger operator \cite{ne98}. 

The consideration of eigenvalue flows of $H$ introduced in \cite{na95} has 
initiated a number of numerical works on such flows, including studies of 
the index theorem \cite{na97}, of topological susceptibility \cite{na98}, 
of instanton effects \cite{ed98}, and of the spectrum gap with the mass 
parameter \cite{edw98}. In these works the derivative of the flows has been 
used \cite{na97,ed98}, considering the respective relation as a result of 
first-order perturbation theory \cite{na97}. In theoretical considerations 
of these flows, as in Section 8 of Ref.~\cite{na95}, details of the 
behavior at crossing have been of interest. 
Further, questions concerning the eigenvalue flows have also been raised 
in investigations \cite{ch98,ad99} of the vicinities of the parameter 
values $-m/r=0,2,4,6,8$.

Generally one should be aware of the fact that the considerations of 
eigenvalue flows of $H$ actually imply certain smoothness conditions to hold, 
which goes beyond the solving of the eigenvalue equation at individual points. 
Locally this requires adequate properties of derivatives and globally that 
integration provides appropriate solutions. In the case of the specific 
hermitean operator considered, in unitary space fortunately there are 
theorems \cite{ka66} which we can use to settle the first point. 
To clarify the second point we have to develop an appropriate procedure 
of integration.

In the present paper we derive a differential equation for the eigenvalues
of the hermitean Wilson-Dirac operator $H$ and give a complete specification
of its admissible solutions. We are able to do this in a mathematically well 
defined way.

Our developments appear important for a number of problems. In particular,
there is the ambiguity in the choice of the mass parameter $m$ which
affects the counting of crossings of flows on a finite lattice. So far,
from an upper bound on the gauge field, a bounding function has been derived 
\cite{ad99} which e.g. allows to disentangle physical and doubler regions. 
One can hope that the differential-equation properties found here allow to
extract more detailed information within this respect. In a study of the 
locality \cite{he99} of the Neuberger operator \cite{ne98} a lower 
bound for $H^2$, also relying on the above gauge-field bound, has been 
important. In such investigations again the differential equation may help 
to get sharper results. In present Monte-Carlo simulations with massless 
quarks (with overlap as well as with domain wall fermions) there are severe 
problems due to too small values occurring for $H^2$ \cite{ne99}. The
proposals to deal with this include projecting out some subspace of small 
eigenvalues of $H$ \cite{ed99}, constructing forms of $H$ which are better 
behaved around zero \cite{neu99}, and looking for more suitably gauge-field 
actions \cite{ne99}. There are now hopes that the differential equation, 
allowing a more detailed insight, may guide to a way out of these 
difficulties.   Studying flows numerically one generally has to 
interpolate between a finite number of points. It appears possible to 
develop more efficient methods for this which use the general properties 
obtained for the differential-equation solutions.

In the following we first derive the differential equation for the eigenvalue
flows of $H$ (Section 2). We then discuss the mathematical properties involved 
and restrictions due to the eigenequation (Section 3). Next we integrate the 
differential equation and give a complete overview of its admissible solutions 
(Section 4).  Finally we collect some conclusions (Section 5).

\section{Derivation of differential equation} \setcounter{equation}{0}

\hspace{3mm}
The Wilson-Dirac operator $X/a$ is given by
\be
X = \frac{r}{2} \sum_{\mu} \na{\mu}\dg\na{\mu} + m
     + \frac{1}{2} \sum_{\mu} \gamma_{\mu}(\na{\mu}-\na{\mu}\dg) 
\label{DW}
\ee
where $(\na{\mu})_{n'n} = \delta_{n'n} - U_{\mu n} \delta_{n',n+\hat{\mu}}$ 
and $0<r\le 1$. Its property $ X\dg = \ga X \ga $ implies that 
\be
H=\ga X 
\label{H}
\ee
is hermitean. The operator $H$ has the eigenequation
\be
H \phi_l = \alpha_l \phi_l 
\label{egH}
\ee
where $\alpha_l$ is real and the $\phi_l$ form a complete orthonormal 
set in unitary space, as one has on a finite lattice. 

Multiplying \re{egH} by $\phi_l\dg\ga$ one gets 
$\phi_l\dg\ga H \phi_l = \alpha_l \phi_l\dg\ga \phi_l$ and summing this
and its hermitian conjugate one has
$\phi_l\dg \{\ga,H\}\phi_l = 2\alpha_l \phi_l\dg\ga \phi_l$. From this
by inserting \re{H} with \re{DW} one obtains
\be
 \alpha_l\, \phi_l\dg\ga \phi_l = m + g_l(m)
\label{gm}
\ee
where
\be
  g_l(m) = \frac{r}{2} \sum_{\mu} ||\na{\mu}\phi_l||^2  \;.
\label{gm1}
\ee
For $g_l(m)$ using 
$||\na{\mu}\phi_l||\le||(\na{\mu}-\Id)\phi_l||+||\phi_l||=2$ one gets
\be
0\le g_l(m)\le 8r \;.
\label{bg}
\ee
Further, abbreviating $(\di \alpha_l)/(\di m)$ by $\dot{\alpha}_l$, we 
obtain
\be
\frac{\di (\phi_l\dg H \phi_l)}{\di m}=\phi_l\dg \dot{H} \phi_l +
\dot{\phi}_l\dg H \phi_l+\phi_l\dg H \dot{\phi}_l =
\phi_l\dg\ga \phi_l+\alpha_l \frac{\di (\phi_l\dg \phi_l)}{\di m}  
\label{dH}
\ee
which means that we have
\be
\dot{\alpha}_l=\phi_l\dg\ga \phi_l \; .
\label{pm}
\ee
Combining \re{gm} and \re{pm} we get the differential equation
\be
\dot{\alpha}_l(m) \alpha_l(m)= m + g_l(m)
\label{dif}
\ee
for the eigenvalue flows of the hermitean Wilson-Dirac operator $H$.

\section{Requirements for solutions} \setcounter{equation}{0}

\hspace{3mm}
In Section 2 actually only continuity of $\phi_l(m)$ would have been needed, 
which can be seen repeating the calculations of \re{dH} in terms of finite 
differences. In Section 4, analyzing properties of solutions, we shall
need $\dot{g}_l(m)$ (at least at certain points) which by \re{gm1} implies
also the existence of $\dot{\phi}_l(m)$. All of this is, however, no problem 
because in the case considered we have derivatives of $\phi_l(m)$ up to any 
order. This follows because for our hermitean operator of form $H(m)=H(0)+m\ga$
in unitary space theorems \cite{ka66} apply by which $\phi_l(m)$ gets 
holomorphic on the real axis.

Of the (continuously) infinite number of solutions of \re{dif}, specified 
by integration constants, only a discrete finite subset occurs for a given $H$,
the selection depending on $H$. The number of solutions in this subset, being 
the number of eigenvectors of $H$, is simply the dimension of the unitary 
space. 

Because of \re{pm} by the continuity of $\phi_l(m)$ also $\dot{\alpha}_l(m)$ 
must be continuous for all $m$ in order that the respective solution 
$\alpha_l(m)$ of \re{dif} belongs to the admissible subset. Since the 
eigenvectors $\phi_l(m)$ have derivatives up to any order this must also hold 
for the admissible $\alpha_l(m)$. Because of the continuity required for 
admissible $\alpha_l(m)$ only solutions of the differential equation \re{dif} 
are admitted which exist for all $m$.

\section{Solutions of differential equation} \setcounter{equation}{0}

\hspace{3mm}
Instead of \re{dif} we first consider the differential equation 
\be
\dot{\beta}_l(m)= 2(m + g_l(m))
\label{dif1}
\ee
which by inserting $\beta_l(m)=\alpha_l^2(m)$ becomes \re{dif}. Integration
of \re{dif1} readily gives 
\be
\beta_l(m)=\beta_l(m_b)+2\int^m_{m_b}\di m'(m'+g_l(m')) 
\label{dint}
\ee
in which particular solutions are determined by the choices of $m_b$ and
$\beta_l(m_b)$. These choices are restricted here by the fact that one 
actually wants real solutions of \re{dif} which meet the requirements
discussed in Section 3. 

To get an overview of the properties of \re{dint} we note that
\be
\int^{m}_{\hat{m}}\di m'(m'+g_l(m'))\;,
\label{my}
\ee
where $\hat{m}$ is an arbitrarily fixed value, has a minimum at $m=m_y$ if
\be 
m_y+g_l(m_y)=0 \mb{and} \dot{g}_l(m_y)>-1 \; .
\label{mg}
\ee 
Because of
\be
H\rightarrow m\ga \mb{for} |m| \rightarrow \infty 
\label{inf}
\ee
one has $\phi_l(m) \rightarrow \chi_{\pm}$ with $\ga\chi_{\pm}=\pm\chi_{\pm}$
in this limit. Then in \re{gm} one gets  
$\alpha_l\, \phi_l\dg\ga \phi_l \rightarrow (\pm m)(\pm1)=m$ so that
one obtains $g_l(m)\rightarrow 0$ for $|m| \rightarrow \infty$. From this
and $g_l(m)\ge 0$ it follows that there is at least one solution 
$m_y\le 0$ of \re{mg}. In general 
several ones with $m_{y_s}< \ldots< m_{y_1}<m_{y_0}\le 0$ may occur.
If there is only one solution of \re{mg} we choose $m_b=m_y$. If there
are several ones we put $m_b$ equal to the $m_{y_{\nu}}$ related to the
lowest minimum of \re{my} (in case of several degenerate ones picking
arbitrarily one of them). In this way we achieve that 
\be
\int^m_{m_b}\di m'(m'+g_l(m'))\ge 0 \mb{for all} m\;.
\label{pos}
\ee 

In order to get a real solution $\alpha_l(m)$ of \re{dif} which exists for all 
$m$, according to \re{pos} (which takes the value 0 for $m=m_b$) one has to 
choose $\beta_l(m_b)\ge 0$ in \re{dint}. At the points $m_{y_{\nu}}$ 
with minima of \re{dint} where $\beta_l(m_{y_{\nu}})>0$ one then immediately
sees that for the solutions of \re{dif} one has a minimum of 
$\alpha_l(m)=+\sqrt{\beta_l(m)}$ and a maximum of 
$\alpha_l(m)=-\sqrt{\beta_l(m)}$. The points where $\beta_l(m_{y_{\nu}})=0$, 
i.e.~the ones with the lowest minima of \re{dint}, however, need special 
consideration. There, to clarify the details at crossing, one has (1) to 
check whether the derivative $\dot{\alpha}_l(m)$ is finite as is necessary 
in view of \re{pm} and (2) to disentangle the solutions related to different 
signs of the square root properly.

To check under which conditions the derivative $\dot{\alpha}(m)$ remains
finite we note that from \re{dif} one gets
\be
\dot{\alpha}_l^2 = \frac{(m + g_l(m))^2}{\beta_l(m)} 
\ee
showing that in case of $\beta_l(\tilde{m})=0$ for some $\tilde{m}$ one
must also have $\tilde{m}+g_l(\tilde{m})=0$ in order that the derivative
remains finite. With these relations holding at $\tilde{m}$ one obtains
\be
\dot{\alpha}_l^2(m) \rightarrow 1+\dot{g}_l(\tilde{m}) \mb{for} 
  m \rightarrow \tilde{m} \;.
\label{da0}
\ee 

Because at the points with $\beta_l(m_{y_{\nu}})=0$ envisaged above \re{mg} 
is satisfied, by \re{da0} the existence of the derivative 
$\dot{\alpha}_l^2(m_{y_{\nu}})$ at these points with $\alpha_l(m_{y_{\nu}})=0$ 
is guaranteed. Further, since in \re{mg} one has $\dot{g}_l(m_{y_{\nu}})>-1$ 
it follows that $\dot{\alpha}_l^2(m_{y_{\nu}})>0$ there. Thus, because 
$\dot{\alpha}_l(m)$ should be continuous (as pointed out in Section 3), there 
must be a crossing point of two solutions  $\alpha_l(m)$, i.e.~ the solutions 
$\mp\sqrt{\beta_l(m)}$ from below must continue as $\pm\sqrt{\beta_l(m)}$ 
above the zero and have the derivatives
\be
\dot{\alpha}_l(m_{y_{\nu}})=\pm\sqrt{1+\dot{g}_l(m_{y_{\nu}})}\ne 0 
\label{da1}
\ee 
at the crossing point. 

The asymptotic behavior $\beta_l(m) \rightarrow m^2$  for 
$|m|\rightarrow \infty$, which implies $\alpha(m) \rightarrow \pm m$, should 
be obvious from \re{inf}. If desired it can readily be worked out in more 
detail estimating the integral in the form 
\be
\beta_l(m)=\beta_l(m_b)+m^2-m_b^2+2\int^m_{m_b}\di m'g_l(m') 
\label{dint1}
\ee
of \re{dint}, which already works using the bound \re{bg}.

\section{Conclusions} \setcounter{equation}{0}

\hspace{3mm}
Establishing an exact relation for the derivatives of the eigenvalues of the 
hermitean Wilson-Dirac operator $H$ we have derived a differential 
equation for the eigenvalue flows of $H$. Referring to appropriate 
theorems also the mathematical aspects 
have been fully clarified. Unambiguous prescriptions for the selection of 
admissible solutions have been given. Integrating the differential
equation and analyzing the features of its solutions a complete overview 
has been obtained. Our results appear advantageous for future theoretical 
developments as well as for applications in numerical works.

\section*{Acknowledgement}

\hspace{3mm}
I wish to thank Michael M\"uller-Preussker and his group for their warm
hospitality.

\end{document}